\documentclass{IEEEcsmag}

\usepackage[colorlinks,urlcolor=blue,linkcolor=blue,citecolor=blue]{hyperref}
\expandafter\def\expandafter\UrlBreaks\expandafter{\UrlBreaks\do\/\do\*\do\-\do\~\do\'\do\"\do\-}
\usepackage{upmath,color}

\usepackage{glossaries}
\usepackage{bbm}
\usepackage{dsfont}
\usepackage[style=ieee, sorting=none]{biblatex}
\usepackage[english]{babel}

\usepackage{hyperref}
\usepackage{nameref}
\usepackage{bm}
\usepackage{tcolorbox}
\usepackage{framed}
\definecolor{light-gray}{gray}{0.95}
\definecolor{shadecolor}{named}{light-gray}
\usepackage{shadethm}
\usepackage{amsthm}
\usepackage[normalem]{ulem}
\usepackage{cancel}

\usepackage{tabularx}
\usepackage{booktabs}
\usepackage{bm}
\usepackage{adjustbox}

\newtheorem{assumption}{Assumption}

\newacronym{qkd}{QKD}{Quantum Key Distribution}
\newacronym{mdi}{MDI}{Measurement-Device Independent}
\newacronym{cm}{CM}{Complex Measurement}
\newacronym{tn}{TN}{Trusted Node}
\newacronym{tfqkd}{TF-QKD}{Twin Field QKD}
\newacronym{otp}{OTP}{One Time Pad}
\newacronym{aes}{AES}{Advanced Encryption Standard}
\newacronym{des}{DES}{Data Encryption Standard}
\newacronym{pqc}{PQC}{post-quantum cryptography}

\jvol{XX}
\jnum{XX}
\paper{8}
\jmonth{Month}
\jname{Publication Name}
\jtitle{Publication Title}
\pubyear{2021}

\begin{document}


\title{Distance-Security Tradeoffs for Repeaterless End-to-End QKD Networks 
}

\author{Sumit Chaudhary}
\affil{Emmy Noether Group Theoretical Quantum Systems Design, Technical University of Munich, Munich, Germany}

\author{Davide Li Calsi}
\affil{Technical University of Vienna, Vienna, Austria}

\author{JinHyeock Choi}
\affil{Emmy Noether Group Theoretical Quantum Systems Design, Technical University of Munich, Munich, Germany}

\author{Marc Geitz}
\affil{T-Labs, Deutsche Telekom AG, Berlin, Germany}

\author{Janis Nötzel}
\affil{Emmy Noether Group Theoretical Quantum Systems Design, Technical University of Munich, Munich, Germany}

\markboth{THEME/FEATURE/DEPARTMENT}{THEME/FEATURE/DEPARTMENT}
\begin{abstract}\looseness-1\gls{qkd} offers provably secure, information-theoretic key exchange, but in long-distance scenarios without quantum repeaters, \glspl{tn} are commonly employed despite introducing critical security risks. We propose a redundant key management method for \gls{qkd} network that combines \gls{tfqkd} (or \gls{mdi}-\gls{qkd}) with a novel key-routing scheme to eliminate the need for truly trusted \glspl{tn}. Quantum measurements are handled entirely within the network, minimizing end-user hardware requirements. Multiple \gls{qkd} links connect intermediate nodes such that a successful attack requires the collusion of multiple adversarial nodes, greatly enhancing security over the traditional \gls{tn} model. In this contribution, we discuss the tradeoff between security, key rates, and distances supported by the new method. Our analysis reveals that the improved redundant key management system may enable true end-to-end connectivity over several thousand kilometers while maintaining high security standards.
\end{abstract}

\maketitle
\footnotetext{}

\chapteri{C}lassical cryptography, exemplified by RSA, has secured global communication for decades by relying on the computational intractability of problems such as prime factorization. This security model is now threatened by quantum computing
. Shor’s algorithm, in particular, enables polynomial-time factorization, rendering RSA and related schemes vulnerable in the quantum era. To address this threat, two main approaches have emerged: \gls{pqc} and \gls{qkd}. \gls{pqc} develops algorithms resistant to known quantum attacks, often based on hard lattice or code problems, and can be integrated with existing infrastructure with minimal changes. However, \gls{pqc} still offers only computational security. \gls{qkd}, in contrast, provides information-theoretic security which is rooted in the laws of physics and independent from assumptions on computational models
\cite{bennett2014quantum}. This makes \gls{qkd}, in principle, a future-proof solution. Yet, practical deployment remains challenging due to limited distances, low key rates \cite{pirandola2020advances}, high costs, dedicated infrastructure requirements, and vulnerabilities from device imperfections \cite{lo2014secure}.

A major obstacle to large-scale deployment of \gls{qkd} is the lack of practical quantum repeaters \cite{briegel1998quantum}. As a result, current \gls{qkd} networks rely on \gls{tn}s as intermediate relays \cite{chen2021implementation, peev2009secoqc, sasaki2011field, stucki2011long, dynes2019cambridge, wang2014field}. Although effective for extending communication range, this architecture introduces a critical vulnerability: compromising a single trusted node invalidates the security of the entire network. This \textit{trusted-node problem} fundamentally limits end-to-end security.
In parallel, most \gls{qkd} protocols require precise and complex quantum measurements, necessitating bulky and highly stabilized hardware. This \textit{complex measurement problem} prevents integration into compact, user-level devices and remains a barrier even if quantum repeaters become available. Together, the \textit{trusted-node problem} and the \textit{complex measurement problem} constitute key obstacles to practical, end-to-end \gls{qkd} services.

To address the \textit{trusted-node problem} in the absence of repeaters, we propose a redundant key-forwarding scheme in which network nodes and end users perform independent \gls{qkd}. The sender masks the secret using \gls{qkd} keys, which are sequentially combined by intermediate nodes and forwarded to the receiver. The end-to-end key is never revealed to any single node; instead, security is distributed across the network, requiring an adversary to compromise multiple nodes simultaneously. 
This scheme is scalable, compatible with existing optical infrastructure, and resilient in the absence of quantum repeaters. Compared to conventional \gls{tn}-based approaches, it significantly reduces single-point failure risks. In addition, to solve the \textit{complex measurement problem}, all quantum measurement apparatuses are absorbed into the network. 

This paper is organized as follows: we propose the network architecture in section \nameref{sec:qkd network}. 
We present the generalized form of the network in section \nameref{sec_gen_net}. Section \nameref{sec_security} discusses the network resilience for various topologies. All the key results and further work are summarized in section \nameref{sec:conclusions}.

\section{QKD NETWORK}
\label{sec:qkd network}
In the following, we describe our key management method. The central goal of the method is to increase the range of \gls{qkd}, use low-complexity hardware at end-user terminals, and offer a security level comparable to that of established \gls{qkd} systems. We note that this method can be implemented using any \gls{qkd} technology. 

In our analysis, we focus on \gls{tfqkd} and \gls{mdi}-\gls{qkd}, as these protocols employ an untrusted intermediate node to perform quantum measurements. This architecture allows bulky, measurement-related quantum hardware to be deployed within the network, while end users require only compact quantum transmitter units. In addition, \gls{tfqkd} enables longer communication distances compared to conventional BB84-based \gls{qkd}. The proposed network architecture consists of point-to-point \gls{qkd} links at the network edges and \gls{tfqkd} or \gls{mdi}-\gls{qkd} links connecting intermediate nodes to end users.

\begin{figure*}
\centerline{\includegraphics[width=37pc]{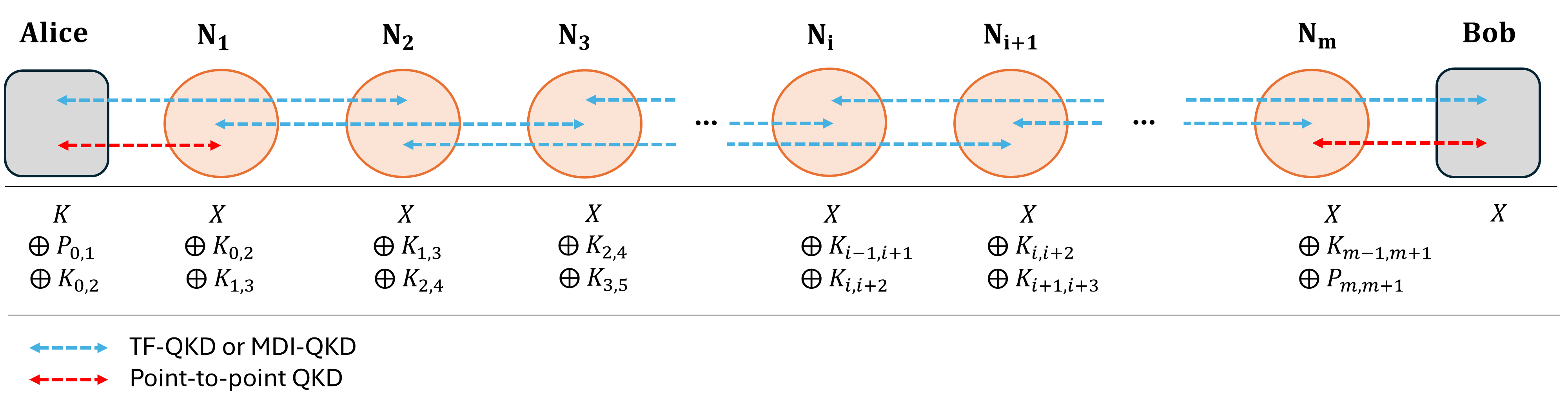}}
\caption{End-to-end QKD network with $m$ intermediaries. At each node, the message obtained through XOR operations on the corresponding keys is depicted below the node and subsequently forwarded to the next node. The blue and red arrows indicate the QKD links, which include all steps such as quantum state preparation, transmission, measurement, and post-processing steps of parameter estimation, error correction, and privacy amplification, resulting in a final secret key. All steps of the key-forwarding mechanism are classical post-processing.}\vspace*{-5pt}
\label{fig_line_2}
\end{figure*}

Let's assume there are $m$ intermediate nodes between Alice and Bob. In this network, every $i^{th}$ node performs \gls{tfqkd} or \gls{mdi}-\gls{qkd} with the $i^{th}+2$ node, including Alice and Bob. Also, Alice with the first node and Bob with the last node perform point-to-point \gls{qkd}. The detail of this network is shown in Fig. \ref{fig_line_2}.

\subsection{Key forwarding mechanism}
\label{sub_key_exchange_procedure}
We first describe the basic mechanism underlying the redundant key management method before generalizing it. Assume Alice and Bob are connected with $m$ intermediary nodes $N_i$ 
for $ 1 \leq i \leq m$ , as in Fig. \ref{fig_line_2}.  
For convenience, we denote Alice as $N_0$ and Bob as $N_{m+1}$. 
For $ 0 \leq i \leq {m-1}$, each $N_i$ is directly connected with $N_{i+1}$ 
and, with its help, can perform \gls{tfqkd} (or \gls{mdi}-\gls{qkd}) with $N_{i+2}$ 
to share a secret key $K_{i, {i+2}}$. 
Take notice that, even though $N_{i+1}$ enables \gls{tfqkd} (or \gls{mdi}-\gls{qkd}),
it does not know $K_{i, {i+2}}$. 
With the following scheme, Alice and Bob can share a secret key without revealing it to the intermediary nodes $N_i$. All the intermediate secret keys are of length $n$ and belong to $\{0,1\}^n$.

\begin{enumerate}
\item Alice (and Bob) perform point-to-point QKD with  $N_1$ (and  $N_m$) 
to share a secret key $P_{0, 1}$ (and $P_{m, {m+1}}$) respectively. 

\item Alice, i.e., $N_0$, runs \gls{tfqkd} (or \gls{mdi}-\gls{qkd}) with $N_2$ to share a secret key $K_{0, 2}$. 
Similarly, for $ 1 \leq i \leq {m-1}$, 
each $N_i$ performs \gls{tfqkd} (or \gls{mdi}-\gls{qkd}) with $N_{i+2}$ to share a secret key $K_{i, {i+2}}$. 
Take notice, now each node has two secret keys 
and at least one of them is not known to 
its immediate neighbor, i.e., the node directly connected to itself. 

\item Alice draws a sequences of $n$ uniformly random bits to generates a secret key  $X \in \{0,1\}^n$. 
Alice then performs XOR operation to $X$ with its secret keys
to produce the message $M_0 = X \oplus P_{0, 1} \oplus K_{0, 2}$. 
Alice sends $M_0$ to the next hop $N_1$. 

\item Take notice that $N_1$ can't retrieve $X$ 
because it doesn't know $K_{0, 2}$. 
Then $N_1$ again performs XOR operation to $M_0$  with its secret keys 
to generate the new message $M_1 = M_0 \oplus P_{0, 1} \oplus K_{1, 3} = X \oplus K_{0, 2} \oplus K_{1, 3}$
and forward it to the next hop $N_2$.
With this operation, $N_1$ decrypts the received message $M_0$ with key $P_{0, 1}$
and encrypt it with $K_{1, 3}$.  

\item For $ 1 \leq i \leq {m-1}$, 
each $N_i$ receives from $N_{i-1}$
the message $M_{i-1} = X \oplus K_{{i-2}, i} \oplus K_{{i-1}, {i+1}}$.  
Then $N_i$ again performs XOR operation to $M_{i-1}$ with its secret keys 
to generate the new message $M_i = M_{i-1} \oplus K_{{i-2}, i} \oplus K_{i, {i+2}} = X \oplus K_{{i-1}, {i+1}} \oplus K_{i, {i+2}}$
and forward it to the next hop $N_{i+1}$.
$N_i$ decrypts the received message with key $K_{{i-2}, i}$
and encrypts it with $K_{i, {i+2}}$.  

\item $N_m$ receives the message $M_{m-1} = X \oplus K_{{m-2}, {m}} \oplus K_{{m-1}, {m+1}}$ from $N_{m-1}$.  
Then $N_m$ again performs XOR operation with its secret keys 
to generate the new message $M_m = X \oplus K_{m-1, m+1} \oplus P_{m, {m+1}}$
and forward it to the next hop $N_{+1}$, i.e. Bob.
As before, $N_m$ decrypts the received message $M_{m-1}$ with key $K_{{m-2}, m}$
and encrypt it with $P_{m, {m+1}}$.  

\item Bob receives $X \oplus K_{{m-1}, {m+1}} \oplus P_{m, m+1}$ from $N_m$. Then Bob, i.e., $N_{m+1}$ can perform XOR operation with its secret keys to decrypts the message to retrieve $X$. Alice and Bob share the secret key $K(A) = X$ for further communication.  
\end{enumerate}

In the above procedure, Alice (and Bob) use their two secret keys for encryption (and decryption) respectively, and all the other intermediary nodes use one for encryption and one for decryption. In each forwarding step, an intermediary node receives a message where $X$ is encrypted with two secret keys. The node knows only one of them, so it can't retrieve $X$. With its two secret keys, the intermediary node removes one of them and adds a new secret key to protect $X$ from the succeeding node. Then the new message is forwarded to the next hop, where the same operation repeats again till the message reaches Bob. 
More formally, one can verify that even if one intermediate node $N_i$ is corrupted by the eavesdropper, all the intermediate messages are encrypted through \gls{otp} with at least one secret key that is unknown to the $N_i$ plus eavesdropper coalition.

\begin{figure*}
\centerline{\includegraphics[width=37pc]{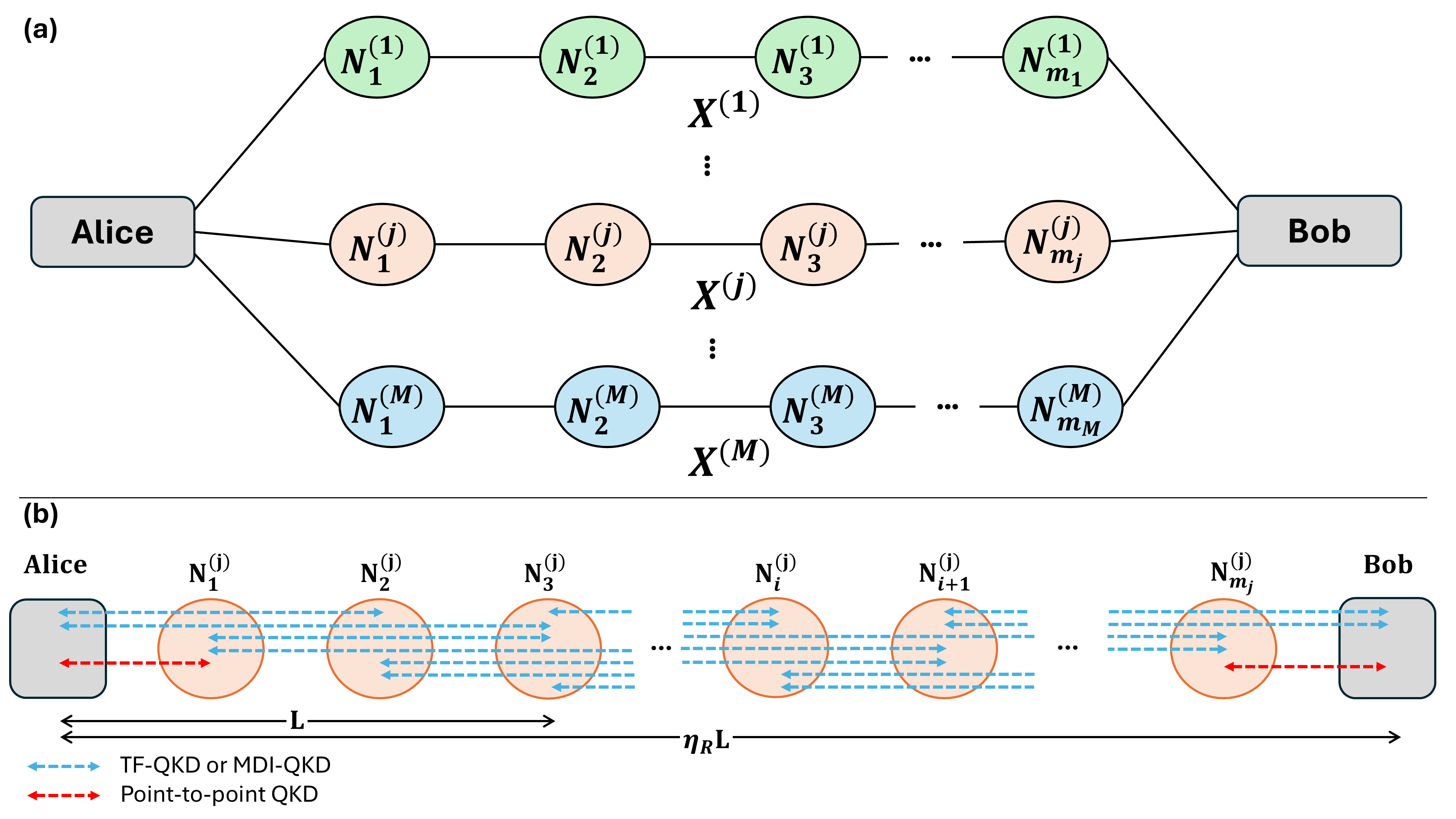}}
\caption{Generalized architecture of the proposed network. (a) Alice and Bob are interconnected through multiple disjoint paths, with each path enabling the exchange of an independent secret bitstring $X^{(j)}$. The final key shared between Alice and Bob is derived by XOR-ing all such bitstrings $X^{(j)}$. (b) Illustration of key distribution among intermediate nodes using \gls{qkd}. In the generalized setting, each node can establish \gls{tfqkd} (or \gls{mdi}-\gls{qkd}) links with up to $t_j$ nearest neighbors (excluding the first nearest node). An example with $t_j = 3$ is depicted. $L$ is the maximum distance between two nodes connected by \gls{tfqkd} (or \gls{mdi}-\gls{qkd}) and $\eta_R L$ is the total length of the network}. \vspace*{-5pt}
\label{fig_multipath}
\end{figure*}

\subsection{Security enhancement}
The resulting scheme is secure unless two adjacent nodes collaborate. 
If $N_i$ and $N_{i+1}$ share their keys, they can easily retrieve $X$. Consequently, the secrecy model is inherently weaker than that of a pure \gls{qkd} link, as only a small fraction of neighboring participants, namely $2/m$, need to collaborate in order to compromise the key. 
However, in the absence of sufficiently advanced quantum hardware capable of implementing quantum repeaters, this represents the most practical method currently available to extend the reach of \gls{qkd}. Furthermore, this is still preferable to the \gls{tn} approach. If Alice and Bob are connected through a chain of trusted \gls{qkd} nodes, every single one of them can successfully cheat on its own. On the other hand, our proposal remains secure against isolated cheaters.
The security of the scheme can be further enhanced by generalizing the protocol to include multiple disjoint paths between Alice and Bob. In this scenario, each path independently transmits a secret key $X_j$ to Bob using the procedure described in \nameref{sub_key_exchange_procedure}. Let's say there are $M$ disjoint paths between Alice and Bob, as shown in Fig. \ref{fig_multipath} (a). Each path is numbered from 1 to $M$, and consists of a sequence $S_j = N_1^{(j)} \dots N_{m_j}^{(j)}$ of $m_j$ independent nodes. Alice and Bob can run $M$ instances of the above scheme, each on one path and with uniformly random input $X_{i} \in_R \{0,1\}^n$. Finally, they can distill a secret key $K(A)=K(B)= \bigoplus_{j=1}^M X^{(j)}$. To leak the key, $M$ pairs of adjacent nodes from each path must collaborate, for a total of $2M$ colluding parties. The availability of such disjoint paths is highly topology-dependent. One can investigate the most common network topologies to investigate realistic values for $M$. In most practical scenarios, we expect $M$ to be at least two. This is because network operators may benefit from having more than a single path between two endpoints, as redundancy in the path number is beneficial for orthogonal reasons (e.g., resilience).

\section{GENERALIZED NETWORK}
\label{sec_gen_net}
The secret key sharing can be performed in a more generalized way by adding more \gls{tfqkd} (or \gls{mdi}-\gls{qkd}) between nodes to tighten the security.
The number of required colluding parties $2M$ stems from the fact that each node shares a secret key with nodes up to distance two, as one can easily verify by inspecting all previous descriptions. One can enhance the security by encrypting with more keys at once, provided that such keys are established with distinct parties. We now proceed to clarify this intuition. \newline
Let us assume that the distance between adjacent nodes is sufficient to perform \gls{qkd}, as before. However, we now add the assumption that the physical links are short enough to perform \gls{tfqkd} (or \gls{mdi}-\gls{qkd}) between $t_j$ nearest nodes. For the $j^{th}$ disjoint path $t_j$ will be positive integer $2\leq t_j \leq m_j$. This network architecture is illustrated in Fig. \ref{fig_multipath}. This assumption enables the aforementioned additional key exchanges. As a result of this highly redundant architecture, the trade-offs between security and hardware investment become a pressing question. We discuss these in the section \nameref{sec_security}, after describing the complete protocol. For the sake of simplicity, we only consider the $j^{th}$ path between Alice and Bob with $m_j$ intermediary nodes. Let us also rename Alice and Bob as $N^{(j)}_0$ and $N^{(j)}_{m_j+1}$ as before.
    \begin{enumerate}
        \item $N^{(j)}_0$ performs point-to-point \gls{qkd} with $N^{(j)}_1$, obtaining key $P^{(j)}_{0,1}$. $N^{(j)}_{m_j+1}$ does the same with $N^{(j)}_{m_j}$, obtaining key $P^{(j)}_{m_j,m_j+1}$.
        \item All the node pairs $N_{i}^{(j)}$ and $N_{i'}^{(j)}$ perform independent \gls{tfqkd} (or \gls{mdi}-\gls{qkd}) such that $2 \leq |i-i'|\leq t_j$, for $i,i' \in \{0,1,2,.. m_j+1 \}$. After all the key exchanges (including the aforementioned point-to-point \gls{qkd}), let $S^{(j)}_i$ denote the set of all bitstring keys known to node $N^{(j)}_i$.
        
        \item $N^{(j)}_0$ draws a uniformly random bitstring $X^{(j)}$ of length $n$ bits. Define $M^{(j)}_{-1} := X^{(j)}$. For $i \in \{0,\dots, m_j\}$, $N^{(j)}_i$ receives message $M^{(j)}_{i-1}$ and computes  $$M^{(j)}_i = M^{(j)}_{i-1}  \oplus \bigoplus_{k \in S^{(j)}_i} k,$$ i.e., it XORs the received message with the XOR of all key bitstrings available at node $N^{(j)}_i$.
        \item $N_{m_j+1}$ receives $M^{(j)}_m$ from $N^{(j)}_{m_j}$, and outputs $$K^{(j)}(B) = M^{(j)}_{m_j} \oplus \bigoplus_{k \in S^{(j)}_{m_j+1}} k$$
    \end{enumerate}
    
The following sequence provides an example for $m_j=4$ and $t_j=3$, showing the intermediate classical messages exchanged during key generation by the parties involved: Alice, Bob, and the intermediate nodes $N^{(j)}_1$, $N^{(j)}_2$, $N^{(j)}_3$ and $N^{(j)}_4$ (the indices 0 and 5 are used for Alice and Bob respectively). Each message $M^{(j)}_k$ is transmitted from the sender to the receiver as described below: \\

\noindent
Alice $\rightarrow$ $N^{(j)}_1$: $M^{(j)}_0 = X^{(j)} \oplus P^{(j)}_{0,1} \oplus K^{(j)}_{0,2} \oplus K^{(j)}_{0,3}$\\ \\
\noindent
$N^{(j)}_1$ $\rightarrow$  $N^{(j)}_2$: $M^{(j)}_1 = X^{(j)} \oplus K^{(j)}_{0,2} \oplus K^{(j)}_{0,3} \oplus K^{(j)}_{1,3} \oplus K^{(j)}_{1,4}$ \\ \\
\noindent
$N^{(j)}_2$ $\rightarrow$  $N^{(j)}_3$: $M^{(j)}_2 = X^{(j)} \oplus K^{(j)}_{0,3} \oplus K^{(j)}_{1,3} \oplus K^{(j)}_{4,1} \oplus K^{(j)}_{2,4} \oplus K^{(j)}_{2,5}$ \\ \\
\noindent
$N^{(j)}_3$ $\rightarrow$ $N^{(j)}_4$: $M^{(j)}_3 = X^{(j)} \oplus K^{(j)}_{1,4} \oplus K^{(j)}_{2,4} \oplus K^{(j)}_{2,5} \oplus K^{(j)}_{3,5}$\\ \\
\noindent
$N^{(j)}_4$ $\rightarrow$ Bob: $M^{(j)}_4 = X^{(j)}  \oplus K^{(j)}_{2,5} \oplus K^{(j)}_{3,5} \oplus P^{(j)}_{4,5}$\\

In the final step, Bob receives a ciphertext corresponding to $X^{(j)}$, which is encrypted using keys known to him (specifically, $K^{(j)}_{2,5}$, $K^{(j)}_{3,5}$, and $P^{(j)}_{4,5}$). Consequently, he possesses sufficient information to correctly decrypt $X^{(j)}$. As the communication range of \gls{tfqkd} (or \gls{mdi}-\gls{qkd}) increases, a larger number of nodes must collude in order to break the protocol. Notably, a single honest node is sufficient to preserve the secrecy of the key. For instance, if $N^{(j)}_3$ is honest but $N^{(j)}_1$ and $N^{(j)}_2$ are malicious, eavesdropping on $M^{(j)}_0, M^{(j)}_1$ and $M^{(j)}_2$ is equivalent to eavesdropping on three copies of $X^{(j)} \oplus K^{(j)}_{0,3}$, which is a \gls{otp} ciphertext obtained by encrypting with secret key $K^{(j)}_{0,3}$. Similarly, eavesdropping on $M^{(j)}_3$ cannot reveal $X^{(j)}$ without knowing $K_{3,5}$ secured by honest node $N^{(j)}_3$. Finally, $M^{(j)}_4$ is encrypted with secret key $P^{(j)}_{4,5}$, and by the properties of \gls{otp} provides no significant information on $X^{(j)}$. Thus, this scheme further distributes trust, thereby improving security.




The proposed protocol can also be equivalently formulated in a matrix-based representation, as detailed in the Supplementary Material. This representation makes explicit that at least \(t_j\) consecutive nodes must collude in order to compromise the secrecy of the $X^{(j)}$. 
Building on this matrix-based analysis, we examine the impact of single \gls{qkd} link failures, noting that the effect on the protocol depends on the position of the failed link. If one of the longest TF/MDI-\gls{qkd} links fails, the effective number of nodes \(t_j\) required to collude and break the protocol for the corresponding disjoint path is reduced by one, while the requirement that compromised nodes must be consecutive remains unchanged to successfully break security. In contrast, if a \gls{qkd} link directly connecting Alice or Bob to the network fails, the effective value of security threshold \(t_j\) and the consecutiveness constraint are relaxed, allowing recovery of the forwarded key \(X^{(j)}\) from \(t_j - 1\) nonconsecutive nodes. For failures of any other single \gls{qkd} link, the protocol remains fully resilient, with neither the required threshold \(t_j\) nor the consecutiveness requirement affected. Based on this analysis, a modified variant of the protocol is also possible, where shorter TF/MDI-\gls{qkd} between network nodes can be totally removed and the protocol will provide a similar security guarantee with less resilience under \gls{qkd} link failure. 

\section{SECURITY ANALYSIS }
\label{sec_security}
We analyze the security of the proposed QKD network by quantifying the probability that an adversary can compromise end-to-end secrecy.
The generalized network topology is characterized by four parameters:
\begin{itemize}
    \item $m_j$: the number of intermediate nodes in the $j$th disjoint path,
    \item $M$: the total number of disjoint parallel paths,
    \item $t_j$: redundancy parameter, i.e., the number of consecutive nodes in the $j^{th}$ path that can be reached in a given direction by each node via \gls{tfqkd} (or \gls{mdi}-\gls{qkd}),
    \item $p$: the probability that an individual node is hacked.
\end{itemize} 

The protocol is considered broken if each disjoint path contains at least $t_j$ consecutive hacked nodes, i.e., a total of $\sum_j t_j$.

\begin{assumption}\label{ass-1}
The hacking probability of any individual node is an independent event, and identical for every network node: the probability that any specific choice of $m'\leq m$ nodes is hacked is given by $p^{m'}$ for some $p\in[0,1]$.
\end{assumption}

Under this assumption, the probability that the protocol is broken, denoted by $P_{\mathrm{break}}$, factorizes over disjoint paths:
\begin{equation}
    P_{\mathrm{break}} = \prod_{j=1}^M \mathcal{P}^{(j)}_{\mathrm{break}}(p,m_j,t_j),
\end{equation}
where $\mathcal{P}^{(j)}_{\mathrm{break}}(p,m_j,t_j)$ is the probability that the $j$th path is compromised.

A path is compromised if it contains at least $t_j$ consecutive hacked nodes. This probability can be expressed as
\begin{equation}
    \mathcal{P}^{(j)}_{\mathrm{break}}(p,m_j,t_j)
    = \sum_{i=t_j}^{m_j} C(m_j,t_j,i)\, p^i (1-p)^{m_j-i},
\end{equation}
where $C(m_j,t_j,i)$ counts the number of configurations in which $i$ hacked nodes include at least $t_j$ consecutive nodes. This formulation captures the inherent robustness of the network: the presence of hacked nodes does not immediately imply a security failure unless they form sufficiently long consecutive segments.

To obtain a tractable upper bound, we replace $C(m_j,t_j,i)$ by the binomial coefficient $\binom{m_j}{i}$ to find out the compact upper bound
\begin{align}
    \mathcal{P}^{(j)}_{\mathrm{break}}(p,m_j,t_j)
    & \leq p^{t_j} (1-p)^{m_j-t_j} 2^{m_j h(t_j/m_j)} \\
    & \leq p^{t_j} 2^{m_j h(t_j/m_j)}
    \equiv \mathcal{P}^{U_j}_{\mathrm{break}},
    \label{eq_p_bound}
\end{align}

where $h(x) = -x\log x - (1-x)\log(1-x)$ is the binary entropy function. A detailed derivation deriving $\mathcal{P}^{U_j}_{\mathrm{break}}$ from
$\mathcal{P}^{(j)}_{\mathrm{break}}(p,m_j,t_j)$ is provided in the supplementary material.
The The term $\mathcal{P}^{U_j}_{\mathrm{break}}$ consists of two competing factors. The factor $p^{t_j}$ provides exponential security as $t_j$ increases. In contrast, the factor $2^{m_j h(t_j/m_j)}$ grows exponentially with $m_j$, although this growth is suppressed when the ratio $t_j/m_j$ approaches either $0$ or $1$. From a network-design perspective, choosing $t_j \approx m_j$ offers little advantage, as it does not extend the effective range of the protocol. On the other hand, very small values of $t_j$ weaken the exponential suppression provided by $p^{t_j}$. Consequently, the parameters $m_j$ and $t_j$ must be carefully balanced to meet the desired security and connectivity requirements of the network.

Considering all disjoint paths, the overall probability that the network is hacked is therefore upper-bounded by
\begin{equation}
    P^U_{\mathrm{break}} = \prod_{j=1}^M \mathcal{P}^{U_j}_{\mathrm{break}}.
    \label{eq_upper_bound}
\end{equation}

Finally, to connect the security analysis to the physical network topology, we assume uniform node spacing along each disjoint path. Let $L$ denote the maximum distance between two nodes connected by TF-QKD (or MDI-QKD), as illustrated in Fig.~\ref{fig_multipath}(b). Defining the range increment factor as $\eta_R = \min_j (m_j + 1)/t_j$, the total end-to-end network range is $\eta_R L$. The total number of intermediate nodes is $N^{(\mathrm{nodes})} = \sum_j m_j$, and the total number of TF-QKD (or MDI-QKD) links required is
\begin{equation}
    N^{(\mathrm{TF/MDI})}
    = \sum_{j=1}^M (t_j - 1)\!\left(m_j - \frac{t_j}{2} + 1\right).
\end{equation}

\begin{figure*}
\centerline{\includegraphics[width=37pc]{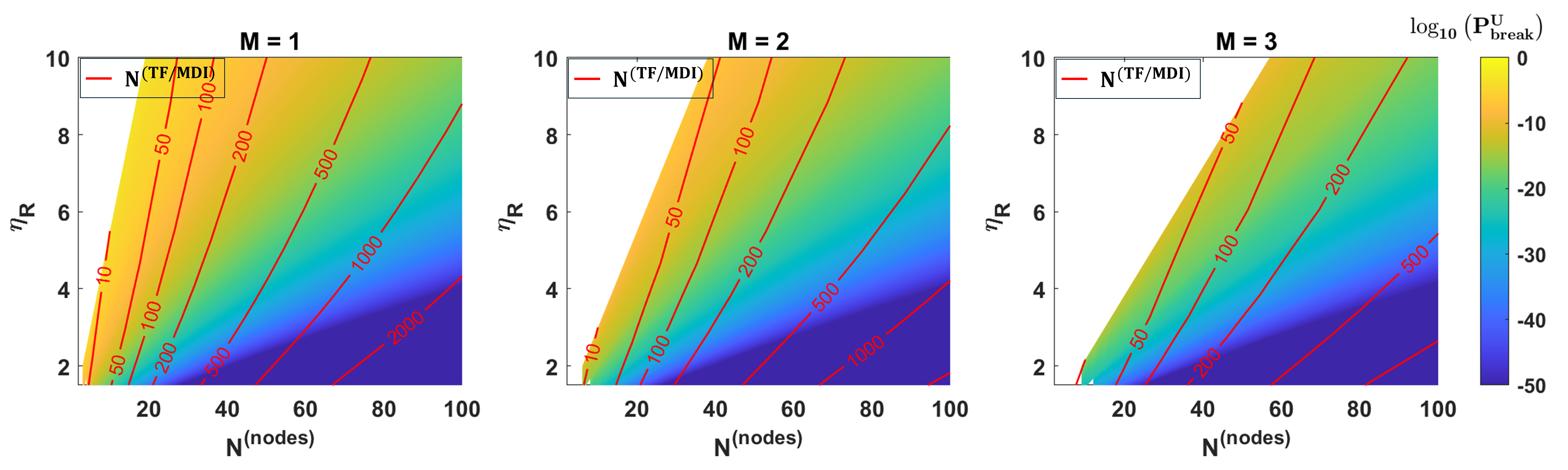}}
    \caption{ The upper bound on the protocol break probability $p_{\text{break}}^U$, is evaluated as a function of the range increment factor $\eta_R$ and the total number of nodes in the network $N^{(nodes)}$. The contour lines indicate the number of \gls{tfqkd} (or \gls{mdi}-\gls{qkd}) links $(N^{(TF/MDI)})$ required to support the network. A node compromise probability of $p=0.001$ is assumed in the calculations. The white regions in the plots correspond to configurations that are not physically realizable, as they violate the physical constraint $2 \leq t \leq m$, which implies $1 + M/N^{(\mathrm{nodes})} \leq \eta_R \leq (N^{(\mathrm{nodes})}/M + 1)/2$.}
    \label{fig_security}
\end{figure*}


\begin{table*}[t]
\centering
\caption{Parameters of end-to-end \gls{qkd} network. ($\bm{N^{\text{(nodes)}}}$ is the total number of intermediate nodes in the network, which does not include Alice and Bob)}
\label{table_qkd_network}

\adjustbox{max width=\textwidth}{
\begin{tabular}{|c|c|c|c|c|c|c|c|c|}
\hline
$\bm{N^{\text{(nodes)}}}$ & $\bm{M}$ & $\bm{m}$ & $\bm{t}$ &
\textbf{Inter-node spacing} & $\bm{N^{\text{(TF-QKD)}}}$ &
\textbf{Network range} & $SKR_{\text{protocol}}$ &
$\bm{P^U_{\text{break}}}$ \\
\hline
32 & 1 (Line topology) & 32 & 4 & 80 km & 93 & 2640 km & 480 bits/s & $1.7 \times 10^{-7}$ \\
32 & 1 (Line topology) & 32 & 6 & 60 km & 150 & 1980 km & 190 bits/s & $5.0 \times 10^{-12}$ \\
32 & 1 (Line topology) & 32 & 4 & 60 km & 93 & 1980 km & 3042 bits/s & $1.7 \times 10^{-7}$ \\
32 & 2 (Disjoint path topology) & 16 & 4 & 80 km & 90 & 1360 km & 480 bits/s & $6.5 \times 10^{-17}$ \\
64 & 2 (Disjoint path topology) & 32 & 4 & 80 km & 186 & 2640 km & 480 bits/s & $2.9 \times 10^{-14}$ \\
\hline
\end{tabular}
}
\end{table*}


\subsection*{Topological Considerations}
We distinguish two categories of topologies:
\begin{enumerate}
    \item \textbf{Line Topology} $(M=1)$: A single sequential path connects Alice and Bob.
    \item \textbf{Disjoint Path Topology} $(M>1)$: Multiple parallel paths connect Alice and Bob.
\end{enumerate}

\begin{figure}
\centerline{\includegraphics[width=18.5pc]{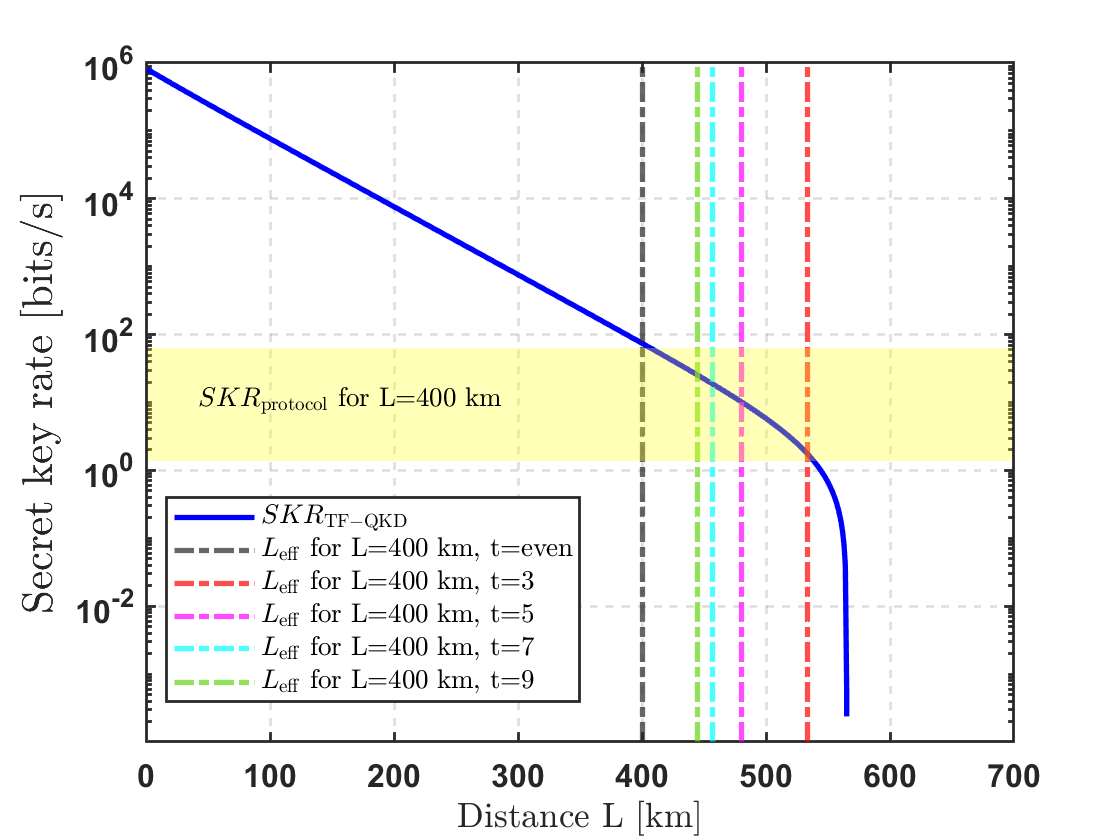}}
    \caption{The blue curve represents the secret key rate $SKR_{\text{TF-QKD}}$ of symmetric TF-QKD as a function of distance. The simulation assumes a fiber loss of 0.2~dB/km, a detector dark count rate of 10~cps, a system repetition rate of 1~GHz, an optical misalignment error of 3\%, and a detector efficiency of 30\%, following Ref.~\cite{lucamarini2018overcoming}. When the longest TF-QKD link in the network is 400~km, the achievable protocol secret key rate $SKR_{\text{protocol}}$ lies within the shaded region, which depends on the redundancy parameter $t$.}
    \label{fig_SKR}
\end{figure}

Security in both topologies is heavily influenced by $t_j$, the required number of consecutive compromised nodes to break a path. Increasing $t_j$ while keeping $m_j$ fixed decreases the likelihood of a security breach exponentially, as the bound $P_{\text{break}}^U \propto \prod_{j=1}^M p^{t_j}$
shows. Consequently, network security can be improved by expanding the network size in terms of increasing the number of nodes or \gls{tfqkd} (or \gls{mdi}-\gls{qkd}) links, as shown in Fig. \ref{fig_security}, depending on the chosen topology and available resources.
Both topologies offer trade-offs between security, node utilization, and range:
\begin{itemize}
    \item The Disjoint Path Topology is well-suited for high-security applications, as it achieves strong security with fewer \gls{tfqkd} (or \gls{mdi}-\gls{qkd}) links $(N^{\text{(TF/MDI)}})$. By distributing nodes across multiple parallel paths, it reduces the number of nodes per path $(m_j)$, which in turn bounds the parameter $t_j\leq m_j$. This configuration minimizes the number of required \gls{tfqkd} (or \gls{mdi}-\gls{qkd}) links, while high security is maintained through the exponential dependence on the number of disjoint paths $M$ in Eq. \ref{eq_upper_bound}. However, the reduced value of $m_j$ also limits the network’s achievable range.
    
    \item Line Topology maximizes communication range by utilizing all available nodes along a single path, resulting in a large $m_j$. To ensure strong security in this configuration, the parameter $t_j$ must also increase, which necessitates a greater number of \gls{tfqkd} (or \gls{mdi}-\gls{qkd}) links. As illustrated in Fig. \ref{fig_security}, this topology can achieve a long-range network even with a relatively small number of nodes $N^{\text{(nodes)}}$, albeit with a trade-off in security. However, this flexibility diminishes as the number of parallel paths $M$ increases. Additionally, the Line Topology benefits from wavelength-division multiplexing, allowing multiple \gls{tfqkd} (or \gls{mdi}-\gls{qkd}) channels to be supported within a single optical fiber.
\end{itemize}

We now analyze the achievable secret key rates for the proposed network parameters. The secret key rate of symmetric TF-QKD (i.e., equal losses in both arms), denoted by $SKR_{\mathrm{TF\text{-}QKD}}$, is shown in Fig. \ref{fig_SKR}.
For simplicity, we assume identical parameters for all disjoint paths, i.e., $m_j = m$ and $t_j = t$ for all $j$. The overall secret key rate of the protocol is determined by the QKD link with the minimum key rate. Since TF-QKD links are longer than point-to-point QKD links, we assume them to be the dominant bottleneck.
In the proposed protocol, the longest TF-QKD link has length $L$ and may be symmetric or asymmetric depending on the value of $t$. For even $t$, both arms are symmetric, whereas for odd $t$, the link becomes asymmetric and the key rate is determined by the arm with higher optical loss. In this case, the effective link length is
\[
L_{\mathrm{eff}} = 2 \left( \frac{L}{t} \left\lfloor \frac{t}{2} + 1 \right\rfloor \right),
\]
where $\lfloor \cdot \rfloor$ denotes the greatest integer function. Consequently, the secret key rate of the protocol $SKR_{\text{protocol}}$ depends only on $L$ and $t$.
Fig. \ref{fig_SKR} presents the achievable secret key rate of the proposed protocol for a longest link length of $L = 400~\mathrm{km}$.

We note that the network design is governed by the choice of parameters $M$, $m$, $t$, and $L$. In contrast, the performance of the proposed network protocol is characterized by the security threshold $P^U_{\text{break}}$, the secret key generation rate $SKR_{\text{protocol}}$, the achievable network range $\eta_R L$, and the required network resources, quantified by the total number of nodes $N^{(\mathrm{nodes})}$ and the number of \gls{tfqkd} (or \gls{mdi}-\gls{qkd}) links $N^{(\mathrm{TF/MDI})}$. Table~\ref{table_qkd_network} illustrates the trade-offs among these parameters for a practical example with $p = 0.001$. The results indicate that enhancing the protocol in terms of security, key generation rate, and network range necessitates increased investment in network resources. Notably, an inter-node spacing of 80~km is consistent with the standard spacing between repeater stations in classical optical communication networks. This similarity enables seamless integration of our model into existing classical communication infrastructure, where pre-installed classical nodes can be upgraded to support \gls{qkd} hardware with minimal structural changes.

\subsection*{Correlated Node Hacking attack}
The assumption of independent node-hacking events provides a tractable and informative baseline for analyzing the protocol; however, in realistic adversarial settings, compromises may exhibit correlations, whereby a successful breach of one node increases the likelihood of subsequent compromises, leading to a cascading effect. We assume that such correlations are confined to directly connected nodes and therefore remain limited to individual disjoint paths. The impact of correlated attacks is analyzed in detail in the Supplementary Material. We summarize the main results below.

\begin{assumption}\label{ass-2}
Consider a disjoint path consisting of $m$ nodes. One node on the path is compromised with probability $p\in[0,1]$. Conditional on this initial compromise, each remaining node on the same path is compromised independently with probability $\Bar{p}\in[0,1]$, where $\Bar{p}>p$. If no initial compromise occurs, no further nodes are compromised.
\end{assumption}

The upper bound on the probability of breaking the $j^{th}$ disjoint path of the protocol is given by:
\begin{equation}
\Bar{\mathcal{P}}^{U_j}_{\text{break}}
=
p \, \Bar{p}^{t_j-1} \, 2^{m_j \, h\!\left(\frac{t_j-1}{m_j}\right)}
\sum_{u=1}^{m_j-t_j+1} (1-p)^{u-1}.
\end{equation}

This $\Bar{\mathcal{P}}^{U_j}_{\text{break}}$ is many magnitude larger than $\mathcal{P}^{U_j}_{\text{break}}$ due to dominant scaling factor $\Bar{p}^{t_j-1}$. We model another correlated attack as follows:
\begin{assumption}\label{ass-3}
The attacker proceeds in a sequential manner. It randomly picks a first node to attack. Given long enough time, it eventually succeeds. Afterwards, it proceeds sequentially, successfully hacking each further node with probability $\hat p$ per pre-defined but arbitrary time slot. A network intrusion detection method is operated in the network which identifies a compromised node with probability $q$.
\end{assumption}
Under assumption \ref{ass-3} the probability for the attacker to hack $t_j$ nodes without getting detected in the $j^{th}$ path is given by
\begin{align}
\textbf{P}(t_j,0)
&= \prod_{i=1}^{t_j}
\frac{\hat{p}(1-q)^i}{1-(1-\hat{p})(1-q)^i} \\
&< \left(\frac{\hat{p}}{q}\right)^{t_j}
(1-q)^{t_j(1+t_j)/2}
\equiv \hat{\mathcal{P}}^{U_j}_{\mathrm{break}} .
\end{align}
As one can see, this probability vanishes exponentially fast in $t_j$ for all values of $\hat{p}$ and $q$ as soon as $t_j$ is large enough.


While correlations can increase the probability of protocol failure, the proposed architecture retains its resilience, and the resulting risk can be mitigated by increasing the number of disjoint paths. This mitigation comes at the cost of additional hardware resources, reflecting an explicit and quantifiable security–resource trade-off inherent in the network design.

The feasibility of deploying the proposed end-to-end \gls{qkd} network is supported by recent field demonstrations and advances in integrated \gls{tfqkd}, underscoring its maturity and scalability. \gls{tfqkd} has been demonstrated over deployed and commercial telecom fibers at distances exceeding 400~km \cite{liu2021field,chen2021twin,pittaluga2025long}, extended beyond 600~km using independent frequency references \cite{zhou2024independent}, and recently realized in chip-based integrated photonic platforms \cite{du2024twin}. Collectively, these results indicate that both the underlying infrastructure and next-generation hardware required for deploying the proposed model are already available, enabling rapid progress toward real-world deployment.

\section{CONCLUSIONS}
\label{sec:conclusions}

In this work, we presented a \gls{qkd} network that adds key improvements to the current \gls{tn}-based network. Our network employs multiple disjoint paths between Alice and Bob, supporting an arbitrary number of intermediate nodes. Unlike the \gls{tn} approach, where a single compromised node can break the entire network, our scheme significantly increases resilience, requiring the collusion of multiple dishonest nodes to breach security. Alternative approaches, such as quantum secret sharing,  exploit multiparty participation to achieve information-theoretic security by reconstructing the secret only when a sufficient fraction of parties are honest; however, these schemes are experimentally more complex and fundamentally constrained by distance. In contrast, our method addresses the long-distance challenge, and its security level can be strengthened to any desired degree by adding more nodes and \gls{tfqkd} (or \gls{mdi}-\gls{qkd}) links.
Our approach is both practical and compatible with near-term \gls{qkd} technologies. It requires only minimal classical communication and postprocessing, with no need for additional quantum hardware.

Furthermore, by placing all complex quantum measurement hardware required for \gls{tfqkd} (or \gls{mdi}-\gls{qkd}) and point-to-point \gls{qkd} within the network, the protocol removes the burden of quantum measurements from Alice and Bob. All measurement-related apparatus is thus absorbed by the network infrastructure. By simultaneously addressing both the \textit{trusted-node problem} and the \textit{complex measurement problem}, this protocol emerges as a strong candidate for scalable, end-to-end \gls{qkd}. Security analysis shows that expanding the network size significantly reduces the likelihood of a protocol breach. The disjoint-path topology provides robust security with fewer \gls{tfqkd} (or \gls{mdi}-\gls{qkd}) links, while the line topology facilitates efficient range extension using a single optical fiber to accommodate all \gls{tfqkd} (or \gls{mdi}-\gls{qkd}) links via wavelength-division multiplexing. As \gls{qkd} technology continues to mature and benefit from economies of scale, this network architecture offers a viable and scalable solution to the \gls{tn} problem. 

Future research directions include investigating the implementation of the proposed network in satellite-based platforms, which offer the potential for significant range improvements due to the nearly lossless space-to-space satellite \gls{qkd} links. Another promising avenue is the analysis of network security under models that account for heterogeneous node compromise probabilities, thereby capturing more realistic adversarial conditions. Furthermore, the framework can be extended to explore alternative network topologies, particularly those supporting multiple end-users, to enhance scalability and broaden practical applicability.

\section{ACKNOWLEDGMENTS}
The authors thank the Q-net-Q Project which has received funding from the European Union’s Digital Europe Programme under grant agreement No 101091732, and is co-funded by the German Federal Ministry of Research, Technology and Space (BMFTR). This work was further financed by the DFG via grant NO 1129/2-1, by the state of Bavaria via the 6GQT project and by the BMFTR via grants 16KISQ077, 16KISQ168, 16KISQ039, 16KIS1598K and	16KISQ093. The authors acknowledge the financial support by the Federal Ministry of Research, Technology and Space of Germany in the programme of “Souverän. Digital. Vernetzt.”. Joint project 6G-life, project identification number: 16KISK002. 
\newline

\textbf{Declaration of competing interests}:
All authors declare no financial or non-financial competing interests.

\def\refname{REFERENCES}

\printbibliography

\begin{IEEEbiography}{Sumit Chaudhary}{\,} is a Ph.D. candidate at the Technical University of Munich, specializing in quantum communication. His research interests include quantum cryptography, optical communication and quantum optics. He received his Master’s degree in Physics from the Indian Institute of Technology Delhi. He can be reached at sumit.chaudhary@tum.de
\end{IEEEbiography}

\begin{IEEEbiography}{Davide Li Calsi}{\,} is a PhD researcher in cryptography at TU Wien, where his work focuses on lattice-based cryptography and pseudorandomness. He received a Master's and Bachelor's degree in Computer Engineering from the Polytechnic University of Milan in 2022 and 2020, respectively.
His contact email address is davide.li.calsi@tuwien.ac.at.
\end{IEEEbiography}

\begin{IEEEbiography}{JinHyeock Choi}{\,} has been involved in research and standard with main focus on network, Internet protocol, complexity, IoT/ CPS and quantum communication. He has contributed to the design, specification and standardization of IP mobility, IPv6 over WiMAX, V2X, Energy efficient network and IoT platform in various organizations such as IETF, WiMAX, IEEE, GreenTouch and OCF. In recent years, he joined Technical Quantum System Design (TQSD) in TUM to work on QKD, quantum token and quantum network. He holds a Ph.D. in Mathematics, Differential Geometry, from POSTECH. He is a member of AMS and has actively participated in Internet standard activities. Contact him at jinchoe@gmail.com.
\vadjust{\vfill\pagebreak}
\end{IEEEbiography}

\begin{IEEEbiography}{Dr. Marc Geitz}{\,} is a researcher at Telekom Innovation Laboratories (T-Labs), Deutsche Telekom AG, Berlin, Germany. His research interests include quantum communication, sensing, networking, and quantum computing applications for industry. He received his Ph.D. in physics from the University of Hamburg in 1999 after studying at RWTH Aachen and working at CERN and DESY. Since 2001, he has worked at Deutsche Telekom. Contact him at marc.geitz@telekom.de \vspace*{8pt}
\end{IEEEbiography}

\begin{IEEEbiography}{Dr. Janis Nötzel} {\,} received the Dipl.-Phys. degree from Technische Universität Berlin in 2007 and the Dr. rer. nat. degree from the Technical University of Munich in 2012. Following a postdoctoral stay at the Universitat Autònoma de Barcelona, he led a research-transfer project at TU Dresden. He currently leads the Emmy Noether Group Theoretical Quantum Systems Design at the Technical University of Munich. His research interests cover quantum technology-based information processing for future networks, spanning information theory, signal processing, software development, and field trials. Contact him at janis.noetzel@tum.de
\end{IEEEbiography}

\end{document}